\documentclass{llncs}
\usepackage{graphicx}
\usepackage{amsmath}

\begin{document}
\title{A numerical trip to social psychology: long-living states of cognitive dissonance}
\titlerunning{A numerical trip to social psychology}  % abbreviated title (for running head)

\author{P. Gawro\'nski\inst{1} \and K. Ku{\l}akowski\inst{1}}
\authorrunning{P. Gawro\'nski et al.}   % abbreviated author list (for running head)
\tocauthor{P. Gawro\'nski, K. Ku{\l}akowski}

\institute{Faculty of Physics and Applied Computer Science, AGH University of Science and Technology, al. Mickiewicza 30, 30-059 Krak\'ow, Poland\\
\email{kulakowski@novell.ftj.agh.edu.pl}}

\maketitle              % typeset the title of the contribution

\begin{abstract}
The Heider theory of cognitive dissonance in social groups, formulated recently in terms of 
differential equations, is generalized here for the case of asymmetric interpersonal ties. 
The space of initial states is penetrated by starting the time evolution several times with 
random initial conditions. Numerical results show the fat-tailed distribution of the time when
the dissonance is removed. For small groups ($N$=3) we found some
characteristic patterns of the long-living states. There, mutual relations of one of the pairs 
differ in sign.
\end{abstract}



\section{Introduction}

The Heider theory of cognitive dissonance in social groups was formulated in 1944 \cite{hei1,hei2}
in terms of relations between triad members. A state is defined as balanced when the following four 
conditions are met: a friend of my friend is my friend, an enemy of my friend is my enemy, 
a friend of my enemy is my enemy, an enemy of my enemy is my friend. In an unbalanced state,
group members suffer from the cognitive dissonance and they try to remove it. It was proven in terms
of the graph theory \cite{har} that a fully connected network is balanced if and only if it is divided into
two antagonistic groups, with all relations between the groups positive and all intergroup relations
negative. The question if the state of balance is ever attained remains open. It is likely
that the answer does depend on the assumed dynamics. Most authors consider the model when
the set of possible states is limited to a positive (friendly) and negative (hostile) one, 
sometimes including zero (neutral or lack of contact) \cite{hum,zth,gir,red1}. In the 
time evolution, these states are changed sharply. In fact, the balanced state is attained 
in all investigated cases, if only the ties were present between all group members 
and they were properly informed on the relations in the group. The case of incomplete information 
was discussed in \cite{hum}.

Recently, the model was reformulated in terms of a continuous change of ties, governed by differential
equations \cite{pg1,pg2,pg3}

\begin{equation}
\frac{dx_{i,j}}{dt}=g(x_{i,j})\sum_kx(i,k)x(k,j)
\end{equation}
where $x_{i,j}(t)$ is the time-dependent relation of $i$ to $j$, and $g(x)$ is a function to bound the relation
$x$ within some prescribed range. Advantages of this modification are that {\it i)} there is no 
ambiguity due to the order of modified ties, and {\it ii)} the condescription is more realistic 
from the psychological point of view. Also, when this approach was applied to some commonly 
discussed examples ("the women of Natchez" \cite{fre} and "the Zachary karate club" \cite{zcr}), the 
results were the same as the best obtained in literature \cite{pg3}. In particular, the division
of the group of 34 club members obtained theoretically was the same as observed by Zachary \cite{zcr,gir}.
Numerical realizations of the system dynamics indicated, that there are two stages of the system. During the first
-- yet unbalanced -- stage the relations vary slowly in an apparently incoherent way. At the end of this stage
they appear to be at the edge of balance, with only some ties to be changed. Then the time evolution 
is accelerated in the sense that the number of unbalanced triads abruptly decreases. Once the balance is attained,
the time derivative of each relation is of the same sign as this relation. Then, the absolute values
of $x_{ij}$ increase till their limits, which depend on the function $g(x)$. 
However, the shape of this function seems not to influence directly the first stage of the balancing process.

Here we generalize the approach to include a possible asymmetry of ties, when the relation of $i$ to 
$j$ is not necessarily the same as the relation of $j$ to $i$. This asymmetry reflects the fact that 
the social relations are never perfectly reciprocated \cite{dkks} ; therefore, our generalization
reflects the actual human behaviour. We ask, if the introduced asymmetry generates any new pattern of 
the system behaviour. In particul1ar we are interested, if imbalanced states can persist. 

\section{The calculations and the results}

Looking for generic solutions, we average the results over a set of initial values of $x_{ij}$ selected randomly
from a narrow range $(-\delta,\delta)$. The same method was applied previously \cite{pg1} for the case of symmetric ties.
These results suggested that in the as-yet imbalanced stage, the values $x_{ij}$ remain small. On the other hand, 
the essential mechanism of removing the cognitive dissonance is captured by Eq. 1 even if $g(x_{ij})$=1 for all ties.
Trying to keep things as simple as possible, we use either $g(x)=1$, or $g(x)=1-(x/R)^2$ with $R$ of the order 
of $10^4$, whereas $\delta =0.5$. Both choices practically neglect $g(x)$ as long as the evolution of $x$ remains 
stationary.

Numerical simulations of the time evolution governed by Eq. 1 are performed for $N$=3,4,5,6 and 7 nodes of 
a fully connected network, within a given 
time period $T=3\times 10^4$. The diagonal elements of the matrix $x_{ii}$ are kept to be zero. Once 
the number of imbalanced triads falls to zero, the calculation is stopped. As for our experience, once the system 
is balanced it remains balanced. The percentage of cases, when 
the balance is not attained within time $T$, is given in Table 1 for various $N$. The total number if simulated 
trajectories is $10^6$ for all values of $N$.
\begin{table}

\begin{center}
        \begin{tabular}{  |c |c| c| }
        \hline \hline
N & $\alpha$ & the percent of unbalanced triads \\ \hline
\hline
3 & -2.07 & 0.08 \\ \hline
4 & -1.83 & 4.21 \\ \hline
5 & -4.4 & 0.43 \\ \hline
6 & -3.7 & 5.43 \\ \hline
7 & $\approx$ -5 & 19.51 \\
\hline
\end{tabular}
\end{center}
\caption{The values of the exponents $\alpha$ and the percentages of trajectories when $\tau >T$ against the number of nodes $N$.}
\end{table}

In Fig. 1 we show the probability distribution of the time $\tau$ when the balance is attained. The obtained curves
indicate that the distribution functions decrease as power functions 
$\rho(\tau)\propto \tau ^{-\alpha}$, with 
the exponent $\alpha$ dependent on the number $N$ of nodes. The values of the exponents 
are given in Table 1. The percentage of the cases when $\tau < T$ oscillates with $N$ but decreases 
for $N=7$; we deduce that the power distribution appears only for small $N$.

\begin{figure}
	\begin{center}
	\includegraphics[angle=-90,width=.8\textwidth]{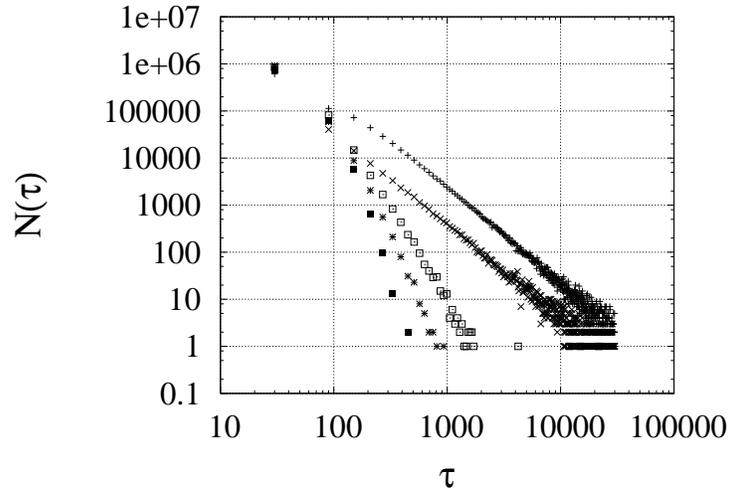}
	\caption{The histograms of the time $\tau$ when the balance is attained, for various 
numbers $N$ of nodes: $N=3,4,6,5,7$ from right to left.}
	\label{fig1}
	\end{center}
\end{figure}

\begin{figure}
	\begin{center}
	\includegraphics[angle=-90,width=.8\textwidth]{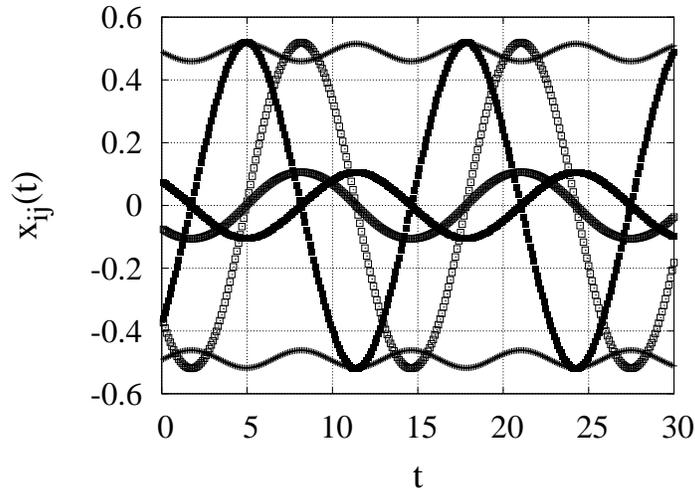}
	\caption{A long-living behaviour of trajectories $x_{ij}$ for $N=3$. There, $x_{23}$ and $x_{32}$ preserve their opposite signs.
Four other $x$'s oscillate, two with the same phase and two with opposite phases.}
	\label{fig2}
	\end{center}
\end{figure}

\begin{figure}
	\begin{center}
	a)\includegraphics[angle=-90,width=.8\textwidth]{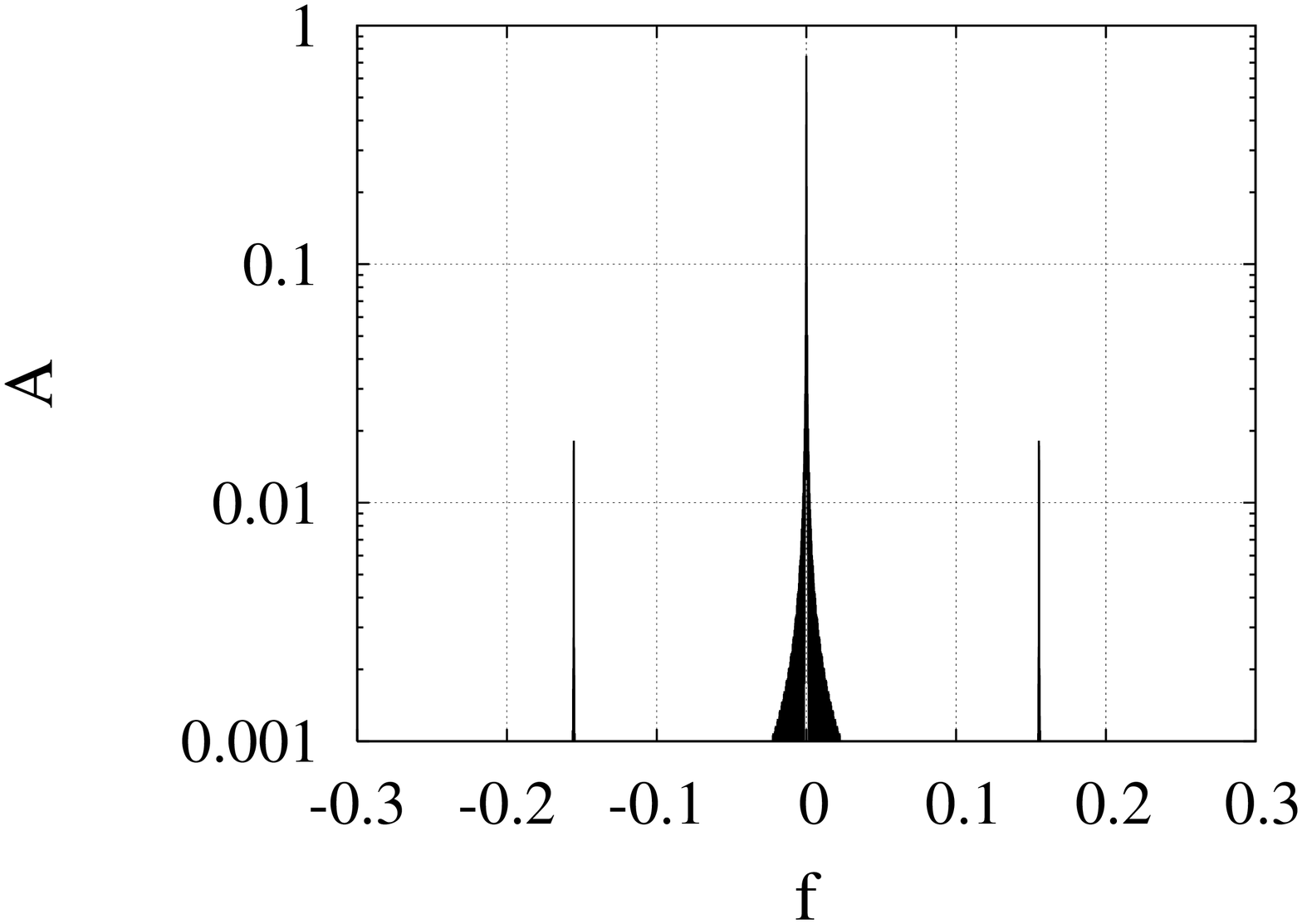}
	b)\includegraphics[angle=-90,width=.8\textwidth]{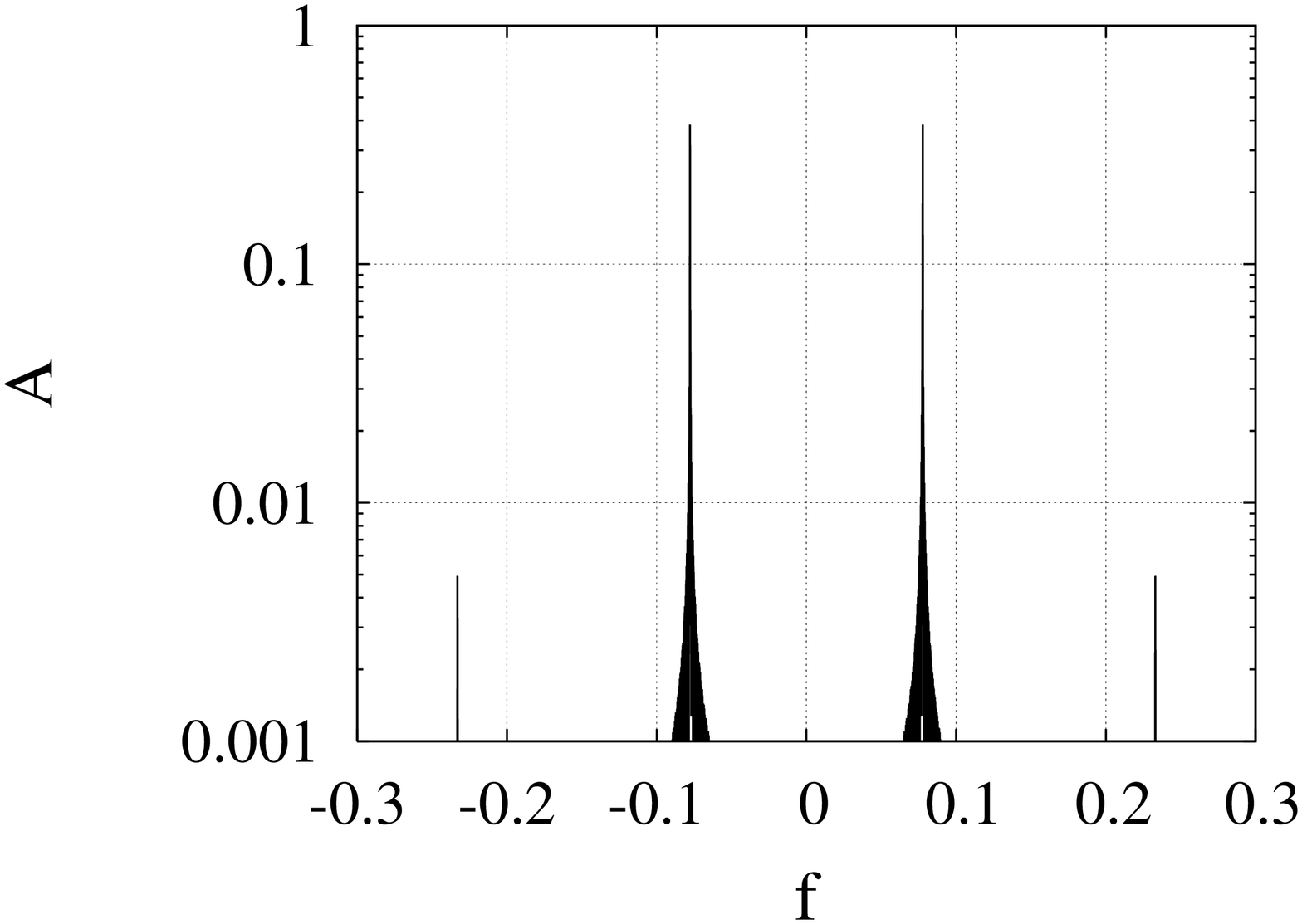}
	\caption{Two Fourier spectra for long-living states when $N=3$; one encountered for $x_{23}$ and for $x_{32}$, and the other --
for the remanining four ties.}
	\label{fig3}
	\end{center}
\end{figure}

\begin{figure}
	\begin{center}
	a)\includegraphics[angle=-90,width=.8\textwidth]{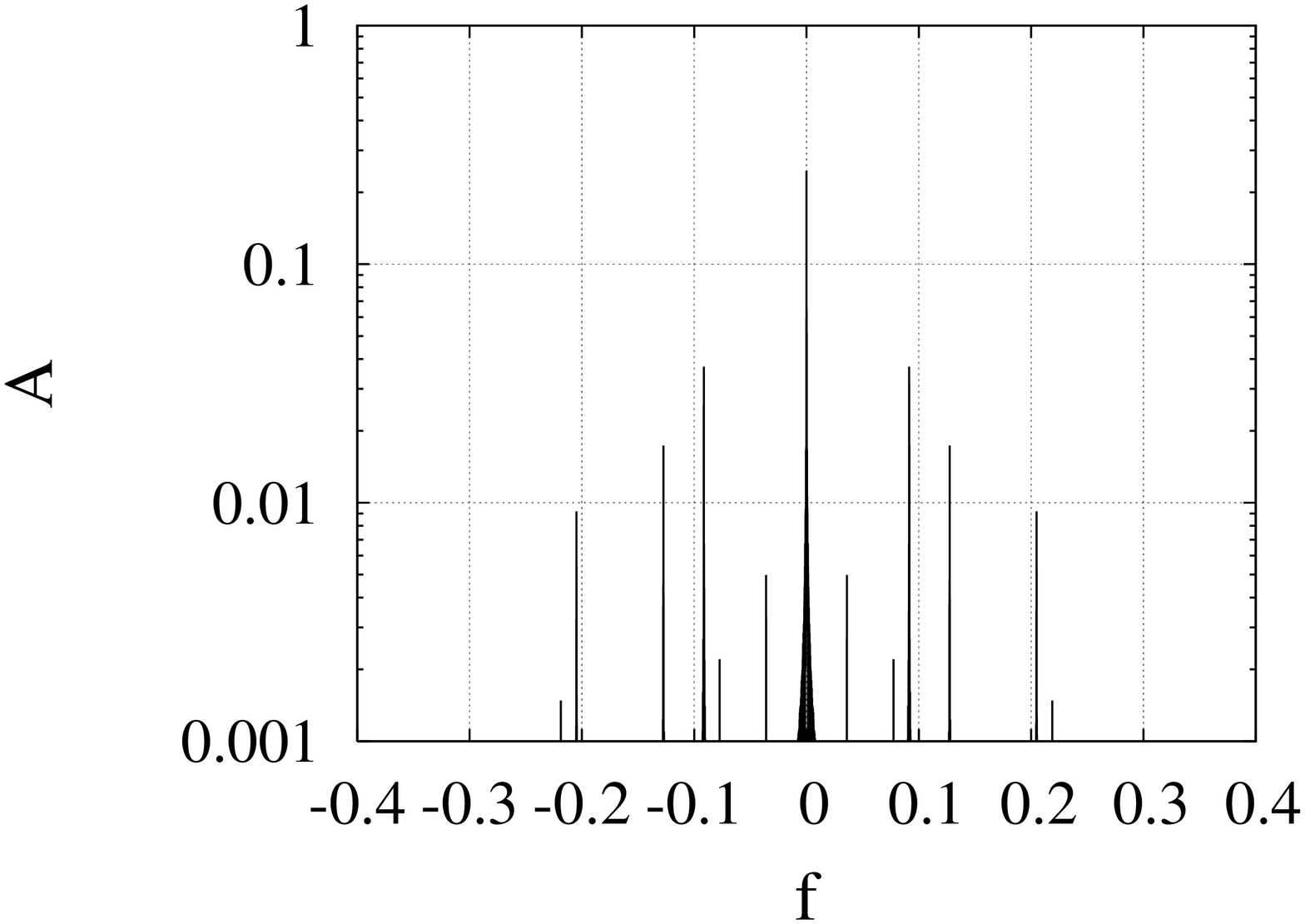}
	b)\includegraphics[angle=-90,width=.8\textwidth]{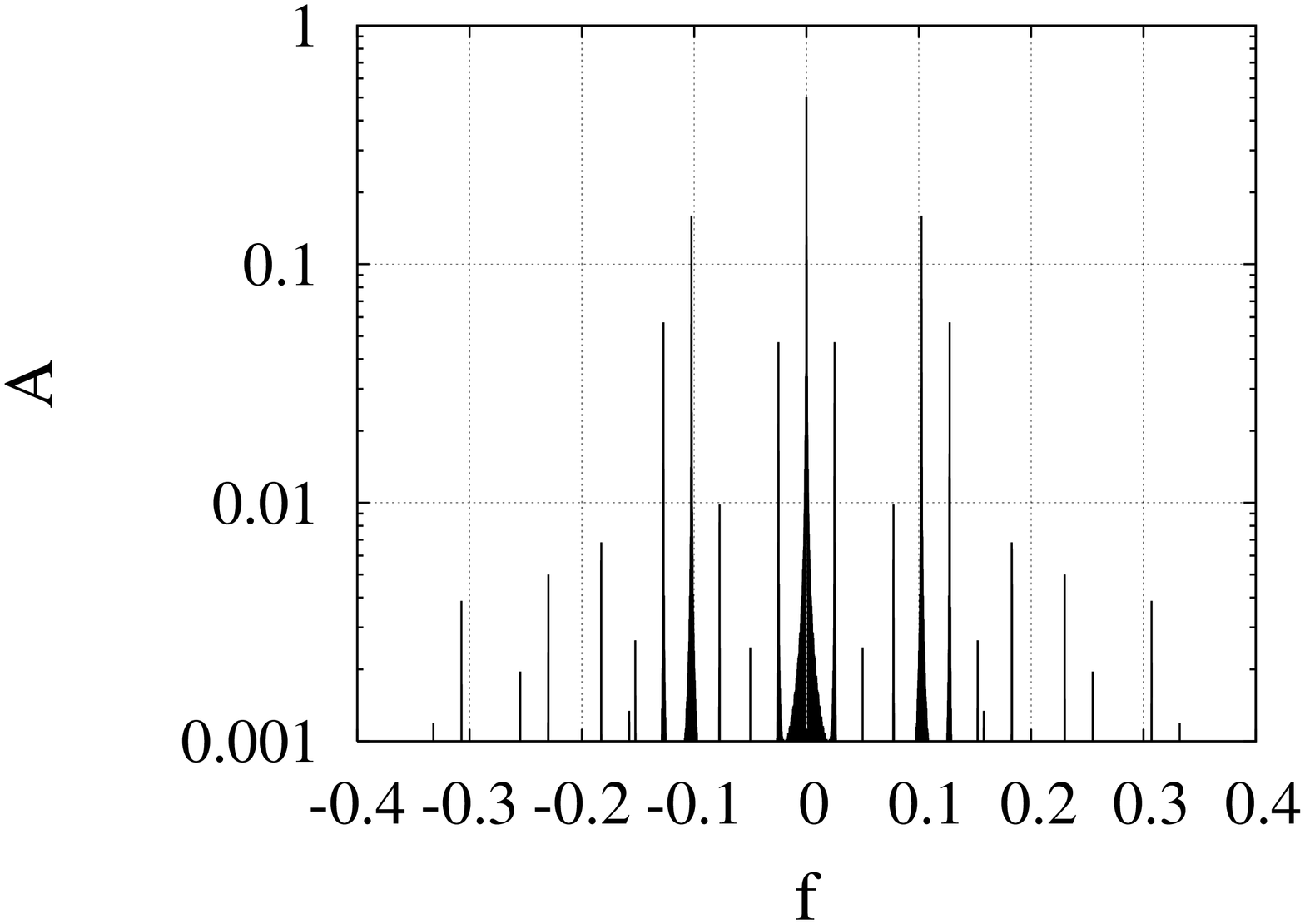}
	c)\includegraphics[angle=-90,width=.8\textwidth]{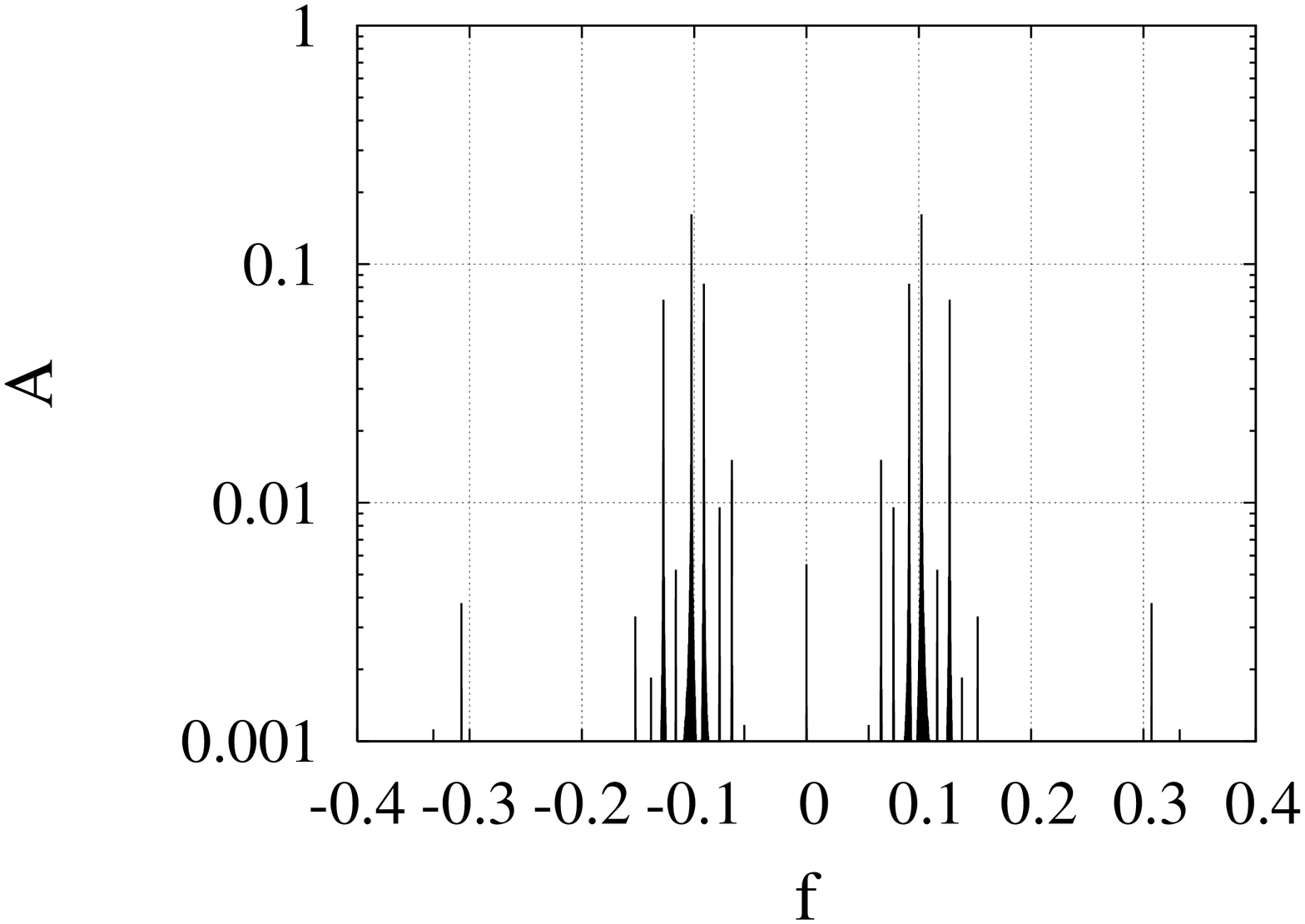}
	\caption{Three Fourier spectra for long-living states when $N=4$; one encountered for $x_{12}$, $x_{21}$, $x_{34}$ and $x_{43}$, 
one for $x_{13}$, $x_{31}$, $x_{24}$ and $x_{42}$, and the other -- for the remanining four ties.}
	\label{fig4}
	\end{center}
\end{figure}

Searching for typical trajectories with large $\tau$, we noticed that indeed some characteristical patterns 
appear which seem stationary or close to stationary. As the number of curves $x_{ij}(t)$ to be investigated depends on $N$
as $N(N-1)$, it is easier to investigate the case when $N=3$. Often, large $\tau$ is produced where the curves
are similar to those in Fig. 2. There is also a regularity in the Fourier transforms of the selected curves.
As a rule, the spectra can be divided in two groups, and those within a group are practically the same. 
Two spectra belong to one group, and four spectra -- to the other. These two belong to the ties which join the same nodes, 
e.g. $x_{12}$ and $x_{21}$. The spectra are shown in Fig. 3. Similar situation is found for $N=4$. In this case, there 
are three different patterns of the Fourier spectra, as shown in Fig. 4. As a rule, ties with similar spectra join 
different nodes, e.g. $x_{12}$ and $x_{34}$ belong to the same pattern.

In the simplest case of $N=3$, these numerical results suggest an approximate analytical solution of Eq. 1. Let us suppose 
that $x_{23}=-x_{32}=\omega$, $x_{ij}=a_{ij}\epsilon \cos(\omega t)$ for the tie $i,j=1,2$ and 
$x_{ij}=a_{ij}\epsilon \sin(\omega t)$ for the tie $i,j=1,3$, where $\epsilon$ is a small parameter. Substituting this 
to Eq. 1, we get that $a_{12}=a_{13}$ and $a_{21}=-a_{31}$. The corrections to $x_{23}$ and $x_{32}$ are of the order 
of $\epsilon ^2$, and to the other $x's$ -- of $\epsilon ^3$. We can deduce that either $sign(a_{12})=sign(a_{21})$ 
and then $sign(a_{13})=-sign(a_{31})$, or $sign(a_{12})=-sign(a_{21})$ 
and then $sign(a_{13})=sign(a_{31})$. The numerical results confirm these sign rules. Also, the Fourier spectra
show that the frequency of the corrections to $x_{23}$ and to $x_{23}$ are characterized by the frequency two times larger 
than the basic frequency characterizing other $x's$. This rule comes directly from the above parametrization, as 
$\dot x_{23}\propto \sin(2\omega t)$.

Some insight is possible also for $N=4$, where the pattern of the 
time evolution is the same for $x_{ij}$ and $x_{ji}$. In this case, all six ties are divided to three pairs, e.g. 
$(1,2)+(3,4)$, $(1,3)+(2,4)$, $(1,4)+(2,3)$. Suppose we denote these sets as $a$, $b$ and $c$. Then the time evolution of 
ties $a$ are governed by the product of elements of $b$ and elements of $c$. Writing it as $\dot a=bc$, we have also 
$\dot b=ca$ and $\dot c=ab$. This regularity is observed in the simulations, and Fig. 4 is typical in this sense.

\section{Discussion}

As follows from Table 1, the exponent $\alpha$ is not universal, as it depends on $N$. However, as can be deduced
from Fig. 1, there is no characteristic scale of time at least for small number of nodes $N$. Translating this result to the social 
psychology, the time $\tau$ of removal the cognitive dissonance can be, at least in principle, arbitrarily long. We note however that
our tool -- the numerical simulation -- does not allow us to state that in some cases, the balance is not attained. On the other
hand, the analysis of the trajectories indicates, that the large values of $\tau$ are due to one ($N=3$) or two ($N=4$)
pairs of ties which are permanently of different sign. This rule could be written symbolically as an old exercise 
for beginneers in Latin: {\it Agricola amat puellam. Puella non amat agricolam} \cite{lat}. The lack of reciprocity seems 
to create the lack of balance; other ties oscillate around zero and therefore they are not able to change the situation.
As a consequence, some stable or metastable patterns appear. Their persistence depends on $N$, and it seems to be 
relatively weak for $N=5$. We deduce it from the fact that in this case the observed times $\tau$ are relatively short. This
can be due to topological properties of the fully connected graph of $N=5$ nodes. The result suggests, that the symmetries
like those discussed above for $N=3,4$ cannot be preserved for $N=5$.

Trying to draw some conclusion for the social psychology, where the original concept of the Heider balance has 
been formulated,
we can refer to some attempts to interpret within the Heider model examples drawn from history or literature. 
In Refs. \cite{har,dor},
a fictitious situation is analysed with four persons: Hero, Blackheart, Buddy and Goodman. In the final balanced state
Blackheart proved to the remaining three that they should act together. The asymmetry which maintained the evolution 
was that Buddy liked Blackheart. Non-reciprocated love as a motif is able to keep the action unsolved and reader's 
attention vivid for long time \cite{crv,cyr}. 
To kill the murderer of his father, Hamlet had to destroy Ophelia's love, as he was not able to implicate her in the conspiracy
\cite{she}. More generally, the unbalanced state is of interest as opposite to an open conflict. It is known that to 
activate enmity, one has to kill a commonly accepted person, as Mohandas Gandhi, 
Martin Luther King 
or Yitzhak Rabin. This method makes the situation clear for warriors.  The case even more fraught with consequences --
 international relationships in Europe from 1872 to 1907-- was mentioned in the context of the Heider balance
in Ref. \cite{ant}. In fact, Otto von Bismarck maintained equilibrium by a set of bilateral relations, binding
hostile states \cite{ndv}. Last but not least, the message {\it Love Thy Enemies} \cite{bbl} can be interpreted as a desperate
attempt to prevent the hate from spreading. These examples suggest that in conflict
 preventing, some asymmetric ties can be essential.


\begin{thebibliography}{00}
\bibitem{hei1} F. Heider, {\it Social perception and phenomenal causality}, Psychol. Rev. {\bf 51} (1944) 358-74.

\bibitem{hei2} F. Heider, {\it The Psychology of Interpersonal Relations}, John Wiley and Sons, New York 1958.

\bibitem{har} F. Harary, R. Z. Norman and D. Cartwright, {\it Structural Models: An Introduction to the Theory 
of Directed Graphs}, John Wiley and Sons, New York 1965.

\bibitem{hum} N. P. Hummon and P. Doreian, {\it Some dynamics of social balance processes: bringing Heider 
back into balance theory}, Social Networks {\bf 25} (2003) 17-49.

\bibitem{zth} Z. Wang and W. Thorngate, {\it Sentiment and social mitosis: implications of Heider's balance theory},
Journal of Artificial Societies and Social Simulation vol. 6, no. 3 (2003) (http://jass.soc.surrey.ac.uk/6/3/2.html)

\bibitem{gir} M. E. J. Newman and M. Girvan, {\it Finding and evaluating community structure in networks}, Phys. Rev. E
{\bf 69} (2004) 026113.

\bibitem{red1} T. Antal, P. L. Krapivsky and S. Redner, {\it Dynamics of social balance of networks}, Phys. Rev. E 
{\bf 72} (2005) 036121.

\bibitem{pg1} K. Ku{\l}akowski, P. Gawro{\'n}ski and P. Gronek, {\it The Heider balance –- a continuous approach},
Int. J. Mod. Phys. C {\bf 16} (2005) 707.

\bibitem{pg2} P. Gawro{\'n}ski, P. Gronek and K. Ku{\l}akowski, {\it The Heider balance and social distance},
Acta Phys. Pol. B {\bf 36} (2005) 2549-58.

\bibitem{pg3} P. Gawro{\'n}ski and K. Ku{\l}akowski, {\it Heider balance in human networks},
AIP Conf. Proc. {\bf 779} (2005) 93-5. 

\bibitem{fre} L. C. Freeman, {\it Finding Social Groups: A Meta-Analysis of the Southern Women Data}, in R. Breiger, 
K. Carley and P. Pattison (eds.): {\it Dynamic Social Network Modeling and Analysis}, The National Academies Press, 
Washington 2003.

\bibitem{zcr} W. W. Zachary, {\it An information flow model for conflict and fission in small groups}, J. Anthropological
Research {\bf 33} (1977) 452-73.

\bibitem{dkks} P. Doreian, R. Kapuscinski, D. Krackhardt and J. Szczypula, {\it A brief history of balance through
time}, J. Math. Sociology {\bf 21} (1996) 113-131.

\bibitem{lat} The farmer likes the girl. The girl does not like the farmer.

\bibitem{dor} P. Doreian and A. Mrvar, {\it A partitioning approach to structural balance}, 
Social Networks {\bf 18} (1996) 149-168.

\bibitem{crv} M. de Cervantes Saavedra, {\it Don Quixote}, London 1885 (http://www.donquixote.com/english.html).

\bibitem{cyr} E. Rostand, {\it Cyrano de Bergerac}, http://www.gutenberg.net/etext/1254.

\bibitem{she} W. Shakespeare, {\it Hamlet, Prince of Denmark}, Oxford UP, London 1914.

\bibitem{ant} T. Antal, P. L. Krapivsky and S. Redner, {\it Social balance of networks: the dynamics of friendship 
and enmity},  presented at Dynamics on Complex Networks and Applications, Dresden, Germany, February 2006 (physics/0605183).

\bibitem{ndv} N. Davies, {\it Europe. A History}, Oxford UP, New York 1996.

\bibitem{bbl} {\it The Bible}, Matt. 5:44.

\end{thebibliography}
\end{document}